\documentclass[%
aip,
rsi,%
amsmath,amssymb,
reprint,%
]{revtex4-1}

\usepackage{graphicx}
\usepackage{dcolumn}
\usepackage{bm}

\begin{document}
	
	\title{Injection locking and coupling the emitters of large VCSEL arrays via diffraction in an external cavity}
	
	\author{Moritz Pflüger}
	\affiliation{Instituto de Física Interdisciplinar y Sistemas Complejos, IFISC (UIB-CSIC), Campus Universitat de les Illes Balears, Ctra. de Valldemossa km. 7.5, 07122 Palma, Spain}

	\author{Daniel Brunner}
	\email{daniel.brunnerfemto-st.fr}
	\affiliation{FEMTO-ST Institute/Optics Department, CNRS \& University Bourgogne Franche-Comt\'e, \\15B avenue des Montboucons,
		Besan\c con Cedex, 25030, France}
	
	\author{Tobias Heuser}%
	\affiliation{Institut für Festkörperphysik, Technische Universität Berlin, Hardenbergstraße 36, 10623 Berlin, Germany}%
	
	\author{James A. Lott}
	\affiliation{Institut für Festkörperphysik, Technische Universität Berlin, Hardenbergstraße 36, 10623 Berlin, Germany}%

	\author{Stephan Reitzenstein}%
	\affiliation{Institut für Festkörperphysik, Technische Universität Berlin, Hardenbergstraße 36, 10623 Berlin, Germany}%
	
	\author{Ingo Fischer}
	\affiliation{Instituto de Física Interdisciplinar y Sistemas Complejos, IFISC (UIB-CSIC), Campus Universitat de les Illes Balears, Ctra. de Valldemossa km. 7.5, 07122 Palma, Spain}

	\date{\today}
	
	\begin{abstract}
		
		Networks of optically coupled semiconductor lasers are of high interest for fundamental investigations and for enabling numerous technological applications in material processing, lighting and information processing.
		Still, experimental realizations of large networks of such elements employing a scalable concepts have so far been been lacking.
		Here, we present a network 22 of the vertical-cavity surface-emitting lasers in a $5 \times 5$ square lattice array via.
		Crucially, the array allows individual control over each laser's pump current, which we leverage spectrally align the array.
		Leveraging diffractive coupling through an external cavity, 22 lasers are mutually injection locked, and, furthermore, we demonstrate their simultaneous phase locking to an external injection laser.
		The VCSEL network is a promising platform for experimental investigations of complex systems and has direct applications as a  photonic neural network.
		The scalability of the concept opens future possibilities for systems comprising many more individual lasers.
		
	\end{abstract}
	
	\maketitle

\section{Introduction}

Semiconductor lasers (SLs) are particularly sensitive to optical coupling and optical feedback, giving rise to a wealth of related phenomena \cite{Wieczorek2005,Soriano2013}.
Various different coupling schemes for lasers have been investigated.
These schemes include integrated photonic circuits \cite{Gao2016,Kao2016,Bao2005}, optical fiber networks \cite{Hill2002,Argyris2016}, topological insulator vertical-cavity laser arrays \cite{Dikopoltsev2021}, and free-space optical setups \cite{Heil2001,Fischer2006,Nixon2012,Brunner2015}.
Applications exploiting the behavior of such (self-) coupled lasers include: 1) phase-locking to increase output powers via coherent beam combining \cite{Gao2016}; 2) secure communications \cite{Koizumi2013,Porte2016,Li2018}; 3) random-bit generation \cite{Uchida2008,Oliver2013,Sakuraba2015,Guo2022}; and 4) brain-inspired computing \cite{Brunner2013,Vatin2019} -- for which possible extensions towards multiple lasers have been proposed \cite{Sugano2020}.

However, experimental realizations of optical coupling for a large number of SLs with significant strength within a scalable approach are still lacking.
Here, we report our advances with vertical-cavity surface-emitting lasers (VCSELs) that are coupled via diffraction in an external cavity \cite{Brunner2015}.
Compared to other SLs, VCSELs stand out due to their low power consumption, nonlinear and fast reaction to optical injection \cite{yu-vcsels,Skalli2022}, the possibility of arranging them in large, practical 2D arrays, and the ability to test them via direct on-wafer probing \cite{yu-vcsels}.
Due to these advantageous properties, VCSELs have been used in various experimental implementations of photonic brain-inspired data processing \cite{Vatin2019,Porte2021,Hejda2021,Skalli2022}.
The diffractive coupling scheme for VCSELs in large arrays reported here is scalable \cite{Maktoobi2019,Bueno2017} and partially reconfigurable \cite{Brunner2015}.
In our particular setup, the VCSELs have been tailored for spectral uniformity and for matched emission polarization. Furthermore, they can be individually addressed via electrical biasing,  allowing us to control the spectral detuning among individual VCSELs \cite{Heuser2020}.

We report on the simultaneous optical injection locking of 22 VCSELs in a $5 \times 5$ square lattice array to an external drive laser.
Without the external drive laser, we achieve mutual locking of 22 VCSELs to a common wavelength and with strongly suppressed autonomous dynamics.
These are promising results in the search for a scalable photonic network platform.

\section{Experimental setup}

For our experiments, we use custom-manufactured GaInAs quantum well VCSELs with an AlGaAs layer for the oxide apertures and AlGaAs/GaAs distributed Bragg reflector (DBR) mirrors to define the central optically $\lambda / 2$-thick cavity \cite{Heuser2020}.
The array used for our experiments consists of VCSELs that are arranged in a $5 \times 5$ square lattice with a pitch of $p \approx \approx{80}{\mu m}$.
They emit with a dominant fundamental transverse mode at $\lambda \approx 976\,\mathrm{nm}$ and exhibit a high degree of homogeneity: spectrally (within $\pm 0.1\,\mathrm{nm}$ at the respective thresholds), in polarization ($\sigma = 4.4^\circ$ due to a slightly elliptical cross-section \cite{Heuser2020}), and regarding their threshold currents $I_\mathrm{th}$ (8\,\% standard deviation when excluding 3 outliers).
Every VCSEL is individually electrically contacted and thus individually addressable, in our case with an 8-bit bias current resolution.
In the following, we will refer to the VCSEL in column $c$ and row $r$ of the array as VCSEL $(c,r)$.

Our diffractive coupling scheme, discussed in detail in \cite{Brunner2015,Bueno2017,Maktoobi2019}, comprises an external cavity, a microscope objective (MO, Olympus LCPLN20XIR, $f_\mathrm{MO}=9\,\mathrm{mm}$, NA\,=\,0.45), a diffractive optical element (DOE, Holoor MS-261-970-Y-X), an achromatic lens (L1, Thorlabs AC254-080-B, $f_\mathrm{AC}=80\,\mathrm{mm}$) and a broadband dielectric mirror (Thorlabs BB1-E03).
The MO collimates the light for passing the DOE, and the lens focuses it on the mirror.
The DOE spatially multiplexes the light.
In our double pass configuration, the $0^\mathrm{th}$ and higher diffractive orders form a $5 \times 5$ pattern, which is imaged back onto the array, where the $0^\mathrm{th}$ order overlaps with the location of the respective source laser and the higher diffractive orders overlap with the location of the corresponding nearest and second-nearest neighboring VCSELs.
This establishes a bidirectional coupling with a strength that decreases with the lattice-distance between the two involved VCSELs.
The reflection at the 50/50 non-polarizing beam splitter cube (BS1, Thorlabs BS014) is used for measurement and analysis, as described further below.

The spatial multiplexing implemented via the DOE also allows for optically injecting an external edge-emitting DBR injection/drive laser diode (Thorlabs DBR976PN) into all 25 VCSELs simultaneously.
This DBR laser is butt-coupled to a polarization-maintaining (PM) single-mode (SM) fiber and an optical isolator (iso).
Its light is collimated using an aspheric lens (L2, Thorlabs AL1225-B, $f=25\,\mathrm{mm}$, NA\,=\,0.23) and polarization-aligned to the VCSELs using a $\lambda /2$ wave plate (Thorlabs WPH10M-980).
This optical drive is injected into the external cavity using the reflection at BS1.
By double-passing the DOE, it is spatially multiplexed similar to the beam of the central VCSEL and thus injected into all the VCSELs simultaneously, although the optical power is not equally distributed.

The VCSEL and DBR laser signal in the analysis branch are split again by a 70R/30T non-polarizing beam splitter cube (BS2, Thorlabs BS023).
The transmitted part is coupled into a SM fiber using an aspheric lens (L3, Thorlabs AL2018, $f=18\,\mathrm{mm}$).
After passing a 50/50 fiber splitter, the light is analyzed using an optical spectrum analyzer (OSA, Anritsu MS9710C, FWHM resolution = 50\,pm) and an optical power meter.
The reflected light is coupled into a multimode (MM) fiber using a plano-convex lens (L4, Thorlabs LA1027-B, $f=35\,\mathrm{mm}$).
The intensity fluctuations are characterized using a photodiode (PD, New Focus 1554-A-50, 10\,kHz to 12\,GHz 3-dB-bandwidth) and an electrical spectrum analyzer (ESA, Anritsu MS2667C, 9\,kHz to 30\,GHz).

\begin{figure}[htb]
	\centering
	\includegraphics[width=1\columnwidth]{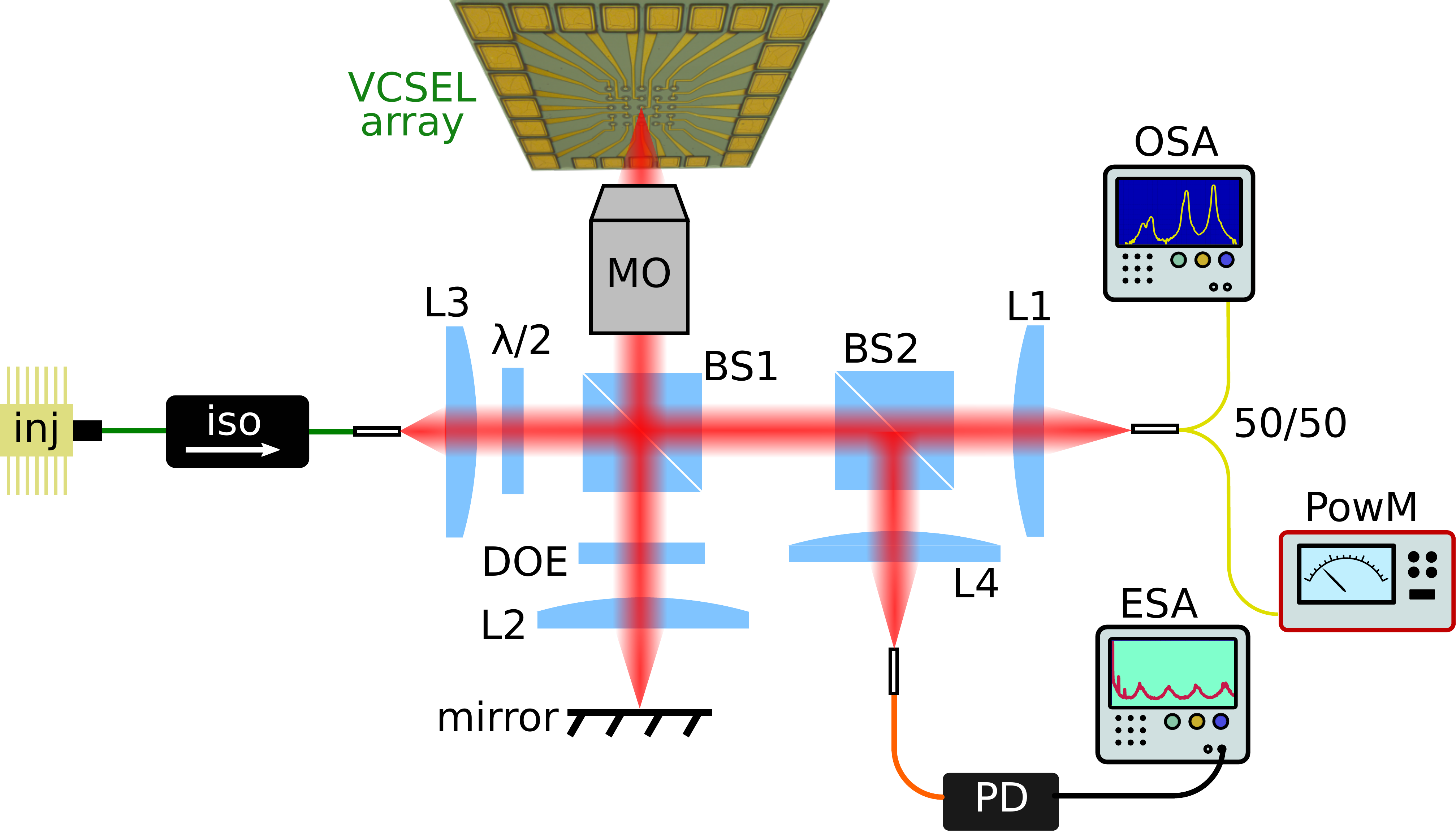}
	\caption{Scheme of the experimental setup. An external cavity is formed by a microscope objective (MO), a lens (L1) and a mirror. A diffractive optical element (DOE) creates multiple beams, establishing coupling between VCSELs and enabling simultaneous optical injection into all the VCSELs of the array. The injection branch consists of a DBR injection laser (inj), an optical isolator (iso), an aspheric lens (L2) and a half-wave plate ($\lambda/2$). Via reflection at a 50/50 beam splitter (BS1), the injected signal enters the external cavity and the VCSEL signal enters the analysis branch, which contains a 70R/30T beam splitter (BS2), an aspheric (L3) and a plano-convex lens (L4), a fiber splitter (50/50), an optical spectrum analyzer (OSA), a powermeter (PowM), a photodiode (PD), and an electrical spectrum analyzer (ESA).}
	\label{fig:exp-setup-simple}
\end{figure}

\section{External optical injection}

\begin{figure*}[htb]
	\centering
	\includegraphics[width=\textwidth]{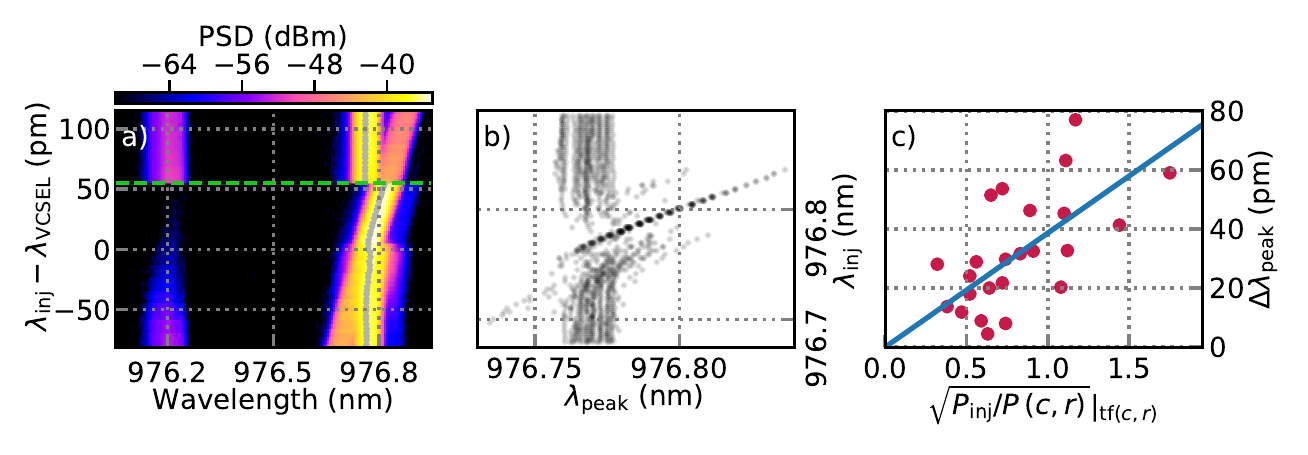}
	\caption{a) Stacked color-coded spectra of VCSEL (2,5) under injection of an external drive laser at different injection laser wavelengths $\lambda_\mathrm{inj}$. Grey trace: Spectral peaks ($\lambda_\mathrm{peak}$), obtained as midpoints between -3\,dB crossings around point of maximal PSD. Dashed green line: $\Delta\lambda_\mathrm{lock}$, i.e. highest value of $\lambda_\mathrm{inj} - \lambda(c,r)$ at which we identify optical injection locking. b) Spectral peaks ($\lambda_\mathrm{peak}$) of 21 injection-locked VCSELs at different $\lambda_\mathrm{inj}$. c) Maximal $\lambda_\mathrm{peak}$ for each VCSEL plotted vs. the expected ratio of injection laser power ($P_\mathrm{inj}$) and VCSEL power ($P(c,r)$) at the surface of VCSEL $(c,r)$. Blue line: Linear fit through zero.}
	\label{fig:osa_inj_3in1}
\end{figure*}

For investigating the VCSELs' behavior under external optical injection, we first spectrally aligned 22 of the 25 VCSELs to $\lambda_\mathrm{VCSEL} = 976.770\,\mathrm{nm} \pm 5\,\mathrm{pm}$.
We compensated for heating of the array caused by electrical biasing.
The remaining spectral inhomogeneity of 5\,pm is due to the 8-bit pump current resolution.
Three short circuits between pairs of VCSELs prevent the exact spectral alignment of more than 22 VCSELs.
For recording the optical spectra shown in this section, a SM fiber and the OSA were positioned behind L4, since the position of the SM fiber behind L3 is needed as an alignment reference and can thus only collect the light of the array's central VCSEL (3,3).
In Fig.\,\ref{fig:osa_inj_3in1}\,a), we show optical spectra of VCSEL (2,5) for different injection laser wavelengths $\lambda_\mathrm{inj}$.
We tuned $\lambda_\mathrm{inj}$ by varying the injection laser temperature $T_\mathrm{inj}$, maintaining its bias current at $I_\mathrm{inj} = 450\,\mathrm{mA}$, which corresponds to an optical output power $P_\mathrm{inj} \approx 17\,\mathrm{mW}$ after the optical isolator, of which $\left. P_\mathrm{inj} \right|_{\mathrm{tf}(2,5)} = {94}{\mu}$ W reach the top facet of VCSEL (2,5).
Due to reflections at the VCSEL's top facet, the emission of the injection laser is also visible in the recorded optical spectrum.

For all but the central VCSEL (due to cross-talk from the injection laser), we observe side-mode suppression and/or a shift of the VCSEL's spectral peak to $\lambda_\mathrm{inj}$.
These are signatures of optical injection locking.
We determine the width of the injection-locking region with two different methods.
First, we analyze $\Delta \lambda_\mathrm{lock} = \lambda_\mathrm{inj} - \lambda_\mathrm{VCSEL}$ at the upper boundary of injection locking.
This is indicated by a dashed green line in Fig.\,\ref{fig:osa_inj_3in1}\,a) and the values are given on the left in Tab.\,\ref{tab:lock-to-inj-det}.
Second, we analyze the wavelength of the peak with the highest power spectral density (PSD) $\lambda_\mathrm{peak}$ as a function of $\lambda_\mathrm{inj}$.
For this, we smoothed the data using a $2^\mathrm{nd}$ order binomial filter and then calculated the midpoint between the -3\,dB crossings at both sides of the peak.
We plot $\lambda_\mathrm{peak}$ as gray dots, c.f. Fig.\,\ref{fig:osa_inj_3in1}\,a), and against $\lambda_\mathrm{inj}$ for all the VCSELs of the array, c.f. Fig.\,\ref{fig:osa_inj_3in1}\,b).
We observe that the slightly spectrally inhomogeneous VCSELs collapse onto $\lambda_\mathrm{inj}$ nearly simultaneously in the region of injection locking.
Note that in some spectra in Fig.\,\ref{fig:osa_inj_3in1}\,b), artifacts arise from the superposition of injection laser and VCSEL spectra.
The maximal values for $\Delta\lambda_\mathrm{peak} = \lambda_\mathrm{peak} - \lambda_\mathrm{VCSEL}$ are given in Tab.\,\ref{tab:lock-to-inj-det} on the right.

From the listed $\Delta \lambda_\mathrm{peak}$, we determine the average coupling efficiency for injection.
For most VCSELs, $\Delta\lambda_\mathrm{peak}$ is close to $\Delta \lambda_\mathrm{lock}$.
This justifies the use of $\Delta \lambda_\mathrm{peak}$ as a measure for the injection locking window, which is given by \cite{Henry1985,ohtsubo}
\begin{equation}
	-\frac{\sqrt{1+\alpha^2}}{\tau_\mathrm{c}}  \left. \sqrt{\frac{K P_\mathrm{inj}}{P(c,r)}} \right|_{\mathrm{tf}(c,r)} \leq \Delta \omega \leq \frac{1}{\tau_\mathrm{c}} \left. \sqrt{\frac{K P_\mathrm{inj}}{P(c,r)}} \right|_{\mathrm{tf}(c,r)},
	\label{eq:lock-range}
\end{equation}
where $\Delta \omega = 2 \pi (f_\mathrm{inj} - f(c,r))$ is the difference of the angular frequencies between drive laser and VCSEL $(c,r)$; $\alpha$, $\tau_\mathrm{c}$, and $P(c,r)$ are the VCSEL's linewidth enhancement factor, its cavity photon lifetime and its output power, respectively; $\left. P_\mathrm{inj} \right|_{\mathrm{tf}(c,r)}$ is the power of the injection laser at the top facet of the VCSEL; and $K$ is the efficiency of coupling into the VCSEL.
This leads to
\begin{equation}
	\Delta \lambda_\mathrm{peak}(c,r) = \frac{\lambda^2}{2 \pi c} \frac{\sqrt{1+\alpha^2}}{\tau_\mathrm{c}} \left. \sqrt{\frac{K P_\mathrm{inj}}{P(c,r)}}\right|_{\mathrm{tf}(c,r)}
	\label{eq:P-frac-prop}
\end{equation}
at the upper injection locking boundary.
We replaced
$\Delta \omega = - 2 \pi c \Delta \lambda / \lambda^2$ with $\Delta \lambda = c / f_\mathrm{inj} - c / f(c,r)$ and $\lambda = c / f_\mathrm{inj} \approx c / f(c,r)$.
$\left. P_\mathrm{inj} \right|_{\mathrm{tf}(c,r)}$ can be expressed as
\begin{equation}
	\left. P_\mathrm{inj} \right|_{\mathrm{tf}(c,r)} = T_\mathrm{MO}  T_\mathrm{BS1}  R_\mathrm{BS1}  C_\mathrm{DOE}(c,r)  P_\mathrm{inj},
	\label{eq:P-inj-tf}
\end{equation}
where $P_\mathrm{inj}$ is the optical power of the injection laser after the optical isolator, $T_\mathrm{MO} = 0.9$, $T_\mathrm{BS1} = 0.48$, and $R_\mathrm{BS1} = 0.46$ are the transmission and reflection coefficients of MO and BS1, respectively, and $C_\mathrm{DOE}(c,r)$ is the DOE's multiplexing matrix coefficient for VCSEL $(c,r)$, which ranges from 1/81 for the corner VCSELs (1,1), (1,5), (5,1), and (5,5) to 1/9 for the central VCSEL (3,3).
In Fig.\,\ref{fig:osa_inj_3in1}\,c), the maximal $\Delta\lambda_\mathrm{peak} = \lambda_\mathrm{peak} - \lambda_\mathrm{VCSEL}$ for each VCSEL $(c,r)$ is plotted versus $\left. \sqrt{P_\mathrm{inj} / P(c,r)} \right|_\mathrm{tf(c,r)}$ with a linear fit, from which we obtain a slope of 39\,pm.
Using Eq.\,(\ref{eq:P-frac-prop}), we obtain $0.6\,\% < K_\mathrm{inj} < 12\,\%$ as an estimate for the coupling efficiency for injection, assuming $2 < \alpha < 5$ and $5\,\mathrm{ps} < \tau_\mathrm{c} < 10\,\mathrm{ps}$.

\begin{table}[htb]
	\centering
	\caption{Right: maximal shift of the spectral maximum due to optical injection locking to the external laser, obtained from -3\,dB crossings ($\Delta\lambda_\mathrm{peak}$). Left: highest detuning between VCSEL and external laser at which locking was observed, extracted from plots like in Fig.\,\ref{fig:osa_inj_3in1}\,a) ($\Delta\lambda_\mathrm{lock}$). We observe good agreement between both sets of values.}
	\label{tab:lock-to-inj-det}
	\begin{tabular}{c|ccccc}
		$\Delta \lambda$ (pm) & (1,*) & (2,*) & (3,*) & (4,*) & (5,*) \\ \hline
		(*,1) & 16/14 & 22/20 & 60/59 & 9/8 & 16/18 \\
		(*,2) & 27/24 & 35/33 & 64/62 & 34/33 & 10/5  \\
		(*,3) & 51/51 & 46/45 & - & 42/41 & 34/32 \\
		(*,4) & 31/29 & 46/46 & 75/76 & 21/20 & 13/9 \\
		(*,5) & 29/28 & 55/53 & 24/22 & 33/33 & 14/12
	\end{tabular}
\end{table}

\begin{figure}[htb]
	\centering
	\includegraphics[width=\columnwidth]{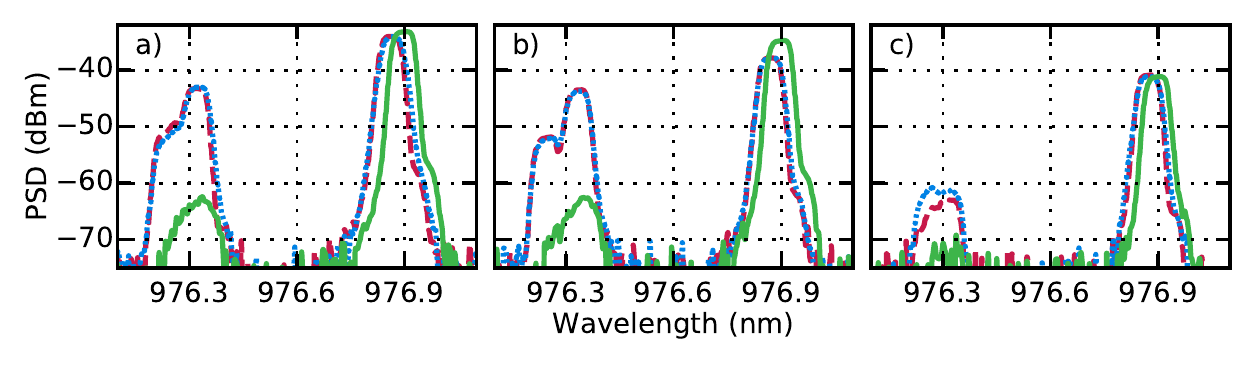}
	\caption{Optical spectra of three different VCSELs in three different configurations each. Dashed red lines correspond to blocking the external cavity, allowing for neither feedback nor injection. Dotted blue lines correspond to blocking the injection path, but not the feedback. Solid green lines correspond to a configuration with both feedback and injection. a) VCSEL (1,4), b) VCSEL (3,5), c) VCSEL (4,5).}
	\label{fig:osa_lock_22}
\end{figure}

Finally, we demonstrate simultaneous injection locking of 22 out of 25 VCSELs in the array.
For this, we first slightly adjusted the pump currents to make side-mode suppression and peak shift more visible.
Then, we recorded optical spectra for every VCSEL in three different configurations, keeping the VCSELs' bias currents and $T_\mathrm{inj}$ constant.
As shown for three examples in Fig.\,\ref{fig:osa_lock_22}, in 21 out of 24 cases, we observed side-mode suppression and/or a shift of the spectral maximum with optical injection, compared to the solitary VCSEL with and without feedback. Since the central VCSEL receives the largest portion of the injected light and is spectrally aligned, we assume that it is injection-locked as well, although due to cross-talk from the injection laser this is impossible to verify.

\section{Pairwise coupling}

\begin{figure}[htb]
	\centering
	\includegraphics[width=\columnwidth]{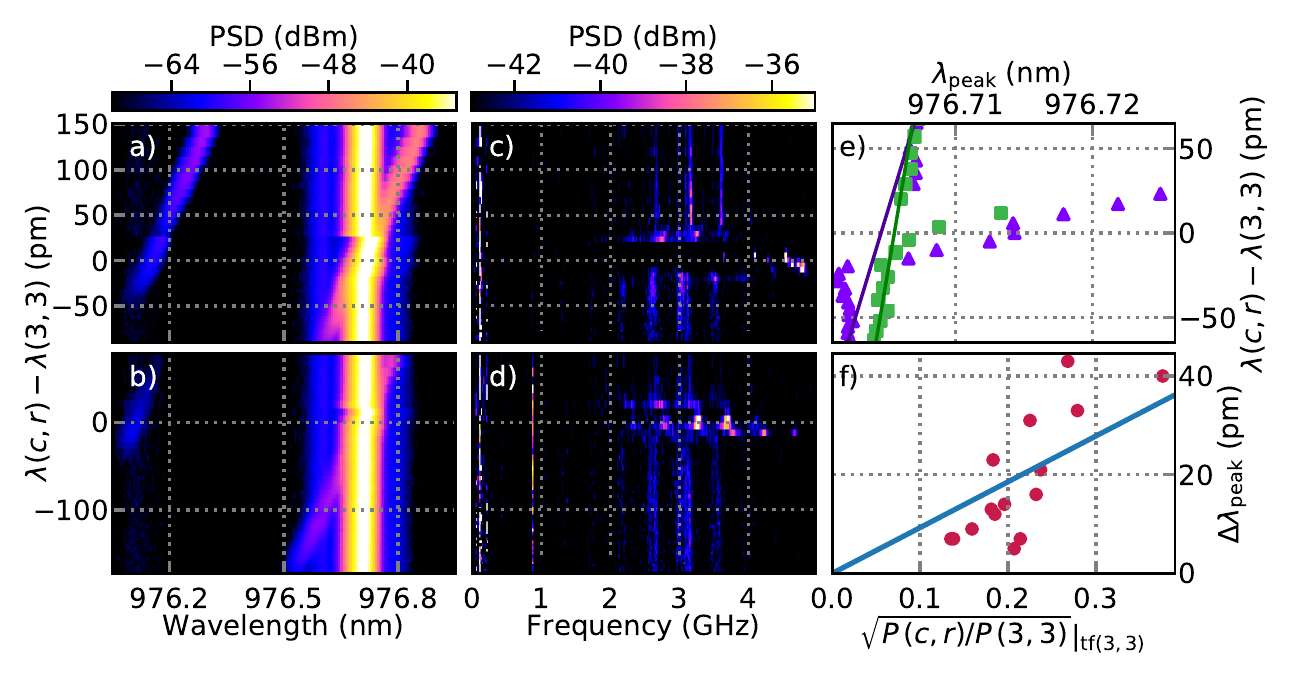}
	\caption{a) Optical spectra of VCSEL (3,3) when tuning the wavelength of VCSEL (3,4), $\lambda(3,4)$. b) Optical spectra of VCSEL (3,3) when tuning $\lambda(2,4)$. Both with the rest of the VCSELs switched off. c) RF spectra corresponding to a). d) RF spectra corresponding to b). e) Peak positions ($\lambda_\mathrm{peak}$) as extracted from a) (violet triangles) and b) (green squares). Lines: linear fits to the points far from the injection locking region. f) Maximal $\lambda_\mathrm{peak}$ observed in the optical spectrum of VCSEL (3,3) for coupling with different VCSELs plotted vs. expected ratio of powers of VCSEL (3,3) and other VCSEL at the surface of VCSEL (3,3). Line: linear fit through zero.}
	\label{fig:osa-pair-6in1}
\end{figure}

We limit our study to pairwise interactions between the central VCSEL (3,3) and one other VCSEL $(c,r)$, as investigating all 231 possible pairwise interactions between independently controllable VCSELs would be unrealistic.
For this, we increased $I(c,r)$ while keeping $I(3,3) = 0.4\,\mathrm{mA} = 1.3\,I_\mathrm{th}$ constant and with all the other VCSELs switched off.
At every step, we recorded the optical and the radio-frequency (RF) spectrum of VCSEL (3,3).
For the pairwise interactions with VCSELs (3,4) and (2,4), these data are shown in Fig.\,\ref{fig:osa-pair-6in1}\,a)\,-\,d), plotted against $\lambda(c,r) - \lambda(3,3)$ at the respective bias currents $I(c,r)$.
Due to reflections, contributions from the second VCSEL are visible in addition to the signal from the central VCSEL.
For more than half of the VCSELs, we observe a shift of the central VCSEL to $\lambda(c,r)$, due to its spectral locking with the other laser \cite{Wunsche2005}.
Furthermore, in the RF spectra, we always observe a signature for pairwise interaction, even for the cases where no clear signature can be found in the optical spectra, see Fig.\,\ref{fig:osa-pair-6in1}\,c)\,and\,d)).
Although these signatures differ for coupling to different VCSELs, in nearly all the cases, at the boundary of the locking region, we observe an increased PSD around the two peaks at 2.6\,GHz and 3.1\,GHz that were previously present with feedback.
There are two more phenomena that we observe in about half of the cases, not necessarily occurring at the same time.
Inside the locking region, we observe a suppression of the two peaks at 2.6\,GHz and 3.1\,GHz that were previously present with feedback, and, at the edge of the locking region, we observe the appearance of multiple peaks at frequencies above 3\,GHz, which are about one external cavity frequency apart.

To quantify the mutual locking, we extract $\lambda_\mathrm{peak}$ from the optical spectra, similar as in the previous section.
For VCSELs (3,4) and (2,4), i.e. for the data from Fig.\,\ref{fig:osa-pair-6in1}\,a) and b), $\lambda_\mathrm{peak}$ is plotted in Fig.\,\ref{fig:osa-pair-6in1}\,e) as a function of the detuning.
Since $\lambda(3,3)$ slightly increases with $I(c,r)$ due to heating of the array, we interpolate $\lambda(3,3)$ by linearly fitting the points far from the locking region.
We then determine the maximum of $\Delta \lambda_\mathrm{peak} = \lambda_\mathrm{peak} - \lambda(3,3)$ for each VCSEL.
Similar to Eq.\,(\ref{eq:P-frac-prop}), we calculate the optical power of VCSEL $(c,r)$ at the top facet of VCSEL (3,3) by
\begin{equation}
	\left. P(c,r) \right|_\mathrm{tf (3,3)} = T_\mathrm{MO}^2  T_\mathrm{BS1}^2  C_\mathrm{DOE}(c,r)  P(c,r).
	\label{eq:P-frac-approx}
\end{equation}
In Fig.\,\ref{fig:osa-pair-6in1}\,f), the maximal $\Delta \lambda_\mathrm{peak}$ for each VCSEL is plotted against $ \left. \sqrt{P(c,r) / P(3,3)} \right|_\mathrm{tf (3,3)}$.
Again, we linearly fit the data and obtain a slope of 32\,pm.
Thus, we arrive at $0.4\,\% < K_\mathrm{pair} < 8.0\,\%$ as an estimate for the pairwise coupling efficiency, which is about 1.5 times lower than for injection from the external drive laser.
Notably, the precision of these measurements is limited by the 50\,pm FWHM resolution of the OSA.

\section{Entire array coupling}

\begin{figure}[htb]
	\centering
	\includegraphics[width=\columnwidth]{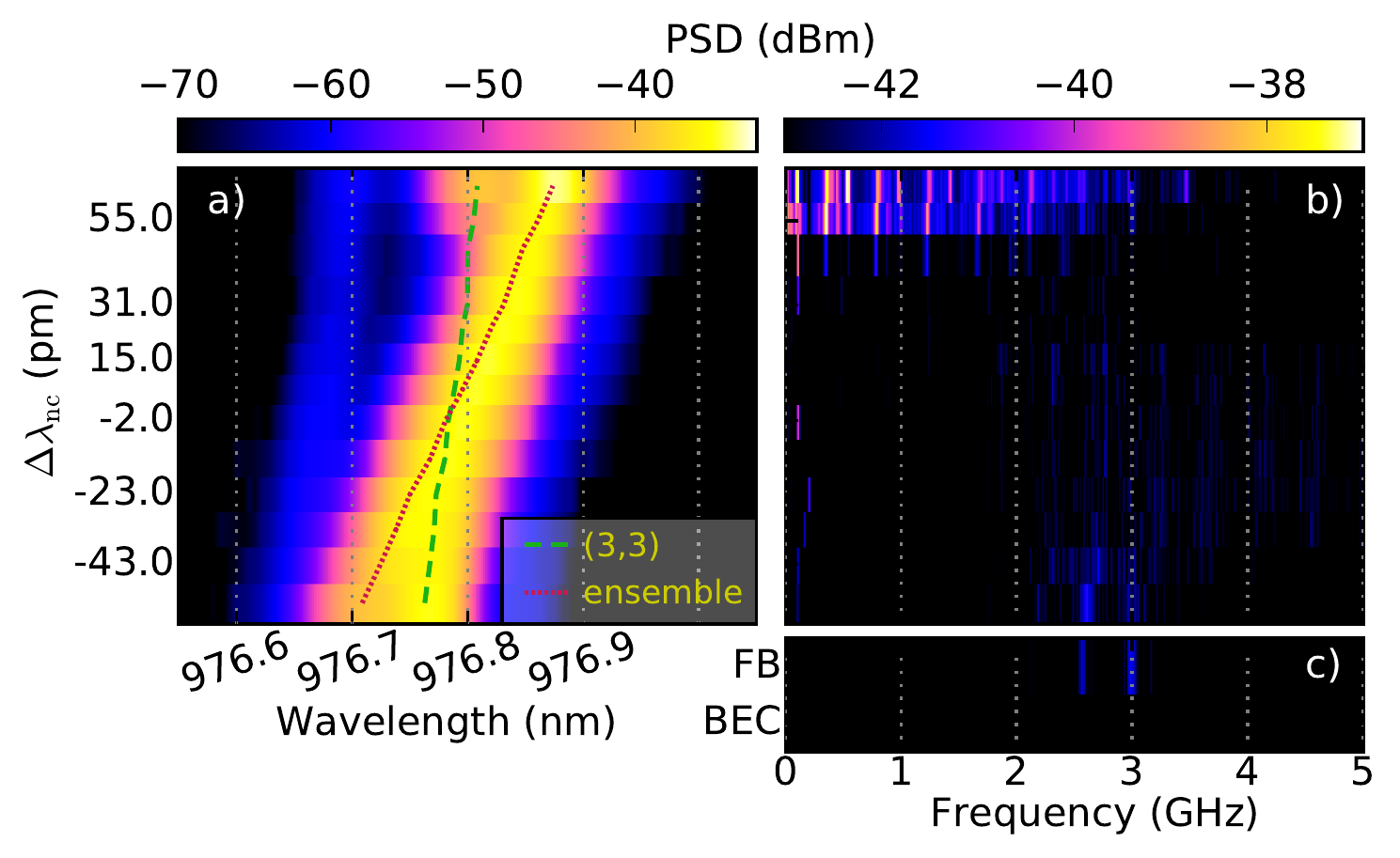}
	\caption{Color-coded spectra of VCSEL (3,3) at $I(3,3)=0.4\,\mathrm{mA}=1.3\,I_\mathrm{th}$ coupled to the entire array. The y-axis represents the spectral detuning $\Delta \lambda_\mathrm{nc} = \lambda_\mathrm{nc} - \lambda(3,3)$. a) Optical spectra. Red dotted line: spectral maximum $\lambda_\mathrm{nc}$ of the ensemble of the non-central VCSELs. Green dashed line: $\lambda(3,3)$. The near-vertical blue trace at about 976.7\,nm stems from VCSEL (3,1), which is short-circuited with VCSEL (3,3) and can thus not be tuned independently. b) Corresponding RF spectra. c) RF spectra of solitary VCSEL (3,3) with feedback ('FB') and blocked external cavity ('BEC').}
	\label{fig:entire}
\end{figure}

To investigate the mutual coupling of the entire array, we kept $I(3,3) = 0.4\,\mathrm{mA}$ constant and simultaneously tuned the pump current of all non-central VCSELs, keeping the tuned VCSELs maximally spectrally homogeneous.
At every step, we recorded an optical and an RF spectrum of VCSEL (3,3), see Fig.\,\ref{fig:entire}.
Due to reflections at the top facet of VCSEL (3,3), contributions of the other VCSELs are visible in the spectra.
To minimize cross-talk due to array heating, we experimentally determined $\lambda(3,3)$ at every step by temporarily blocking the external cavity and recording the optical spectra, thus avoiding reflections and optical injection locking effects.
We also experimentally determined $\lambda_\mathrm{nc}$, the spectral peak of the ensemble of the non-central VCSELs, by recording spectra with VCSEL (3,3) switched off, i.e. the summed spectra of all the non-central VCSELs' emissions through their reflection at the top facet of VCSEL (3,3).

We observe a clear transition in both optical and RF spectra upon increasing $\Delta\lambda_\mathrm{nc} = \lambda_\mathrm{nc} - \lambda(3,3)$.
Two more datasets showing similar behavior have been recorded.
In Fig.\,\ref{fig:entire}, the main transition occurs between $\Delta\lambda_\mathrm{nc} = 47\,\mathrm{pm}$ and $\Delta\lambda_\mathrm{nc} = 55\,\mathrm{pm}$.
In the optical spectra, for $\Delta\lambda_\mathrm{nc} \leq 47\,\mathrm{pm}$, VCSEL (3,3) is optically locked with the rest of the VCSELs before the previously suppressed peak at $\lambda(3,3)$ reappears for $\Delta\lambda_\mathrm{nc} \geq 55\,\mathrm{pm}$.
In the RF spectra, peaks and an increased floor appear at frequencies below 3\,GHz for $\Delta\lambda_\mathrm{nc} \geq 55\,\mathrm{pm}$.
We expect residual wavelength inhomogeneities between different VCSELs due to the 8-bit pump current resolution to be up to 0.01\,nm, which corresponds to up to 3\,GHz.
Thus, we interpret the peaks that appear in the RF spectra for $\Delta\lambda_\mathrm{nc} \geq 55\,\mathrm{pm}$ as beating between different VCSELs.
Since this beating appears at the same time as the unlocking of VCSEL (3,3), we conclude that all 22 independently tunable VCSELs are mutually optically locked for $\Delta\lambda_\mathrm{nc} \leq 47\,\mathrm{pm}$.

Again, we estimate the coupling efficiency using $\Delta\lambda_\mathrm{nc} = 47\,\mathrm{pm}$ as the upper locking boundary, and $\sum_{(c,r)} \left. P(c,r) \right|_\mathrm{tf (3,3)} / P(3,3) \approx 1.14$ as the power ratio, similar to Eq.\,(\ref{eq:P-frac-approx}).
With these values, we obtain a coupling efficiency comparable to $K_\mathrm{inj}$ and higher than $K_\mathrm{pair}$.
Since Eq.\,(\ref{eq:lock-range}) assumes a single monochromatic source and not several VCSELs with uncorrelated phases, one would expect the coupling efficiency for entire array coupling (incomplete locking) to be smaller than the one for pairwise coupling.
That this is not the case corroborates our claim of entire array locking.
Importantly, we do not find the same unlocking characteristics in the RF spectra for negative frequency detuning.
We assign this predominantly to the fact that the emission power of all VCSELs is around 5\,dB lower at the estimated lower boundary of the emission region than at the upper one.
This is comparable to the dynamic range of the RF spectra shown in Fig.\,\ref{fig:entire}.

\section{Conclusion}

In conclusion, we present results on optical injection and diffractive coupling of VCSELs in a 5$\times$5 square lattice array.
For all individual VCSELs, we achieve optical injection locking to an external drive laser and determine a coupling efficiency of $0.6\,\% < K_\mathrm{inj} < 12\,\%$.
Furthermore, we achieve simultaneous optical injection locking of 22 out of the 25 array VCSELs to the external drive laser.
Based on the central VCSEL's RF spectra, we show clear signatures of pairwise coupling between the central and all other VCSELs of the array and demonstrate pairwise optical locking for 13 out of 21 VCSELs/pairs of short-circuited VCSELs.
We estimate a coupling efficiency of $0.4\,\% < K_\mathrm{pair} < 8.0\,\%$, the precision of which is limited by the resolution of the OSA and the low power ratio between the different VCSELs.
When coupling the entire array, we observe a simultaneous transition in both optical and RF spectra of the central VCSEL.
We interpret this as a transition from optical locking of the entire array to unlocking.

Our findings show that custom-engineered VCSEL arrays with external diffractive optical coupling are feasible platforms for realizing large-scale networks of SLs.
Such laser networks show potential for applications in laser machining.
They are also of interest for experimentally studying fundamental properties of complex systems.
Last but not least, they offer attractive properties for optical machine learning.
Our results demonstrate that the obtained coupling between the VCSELs creates a network.
Moreover, optical injection locking to an external drive laser can be achieved, enabling information injection into the system.
Finally, trainable readout weights can be readily implemented using a spatial light modulator \cite{Bueno2017}.
This would create a fully hardware-implemented and parallel photonic neural network with correspondence of one laser per artificial neuron and potentially high-bandwidth operation.	
	
\section*{Funding}
 The authors acknowledge the Volkswagen Foundation for funding within the projects NeuroQNet I\&II. 
M.P. and I.F. acknowledge the Spanish State Research Agency, through the Severo Ochoa and María de Maeztu Program for Centers and Units of Excellence in R\&D (MDM-2017-0711) funded by MCIN/AEI/10.13039/501100011033.

\bibliography{mybib}
\bibliographystyle{ieeetr}
	
\end{document}